\def\be{\begin{equation}}
\def\ee{\end{equation}}
\def\ba{\begin{array}}

\def\ea{\end{array}}

\documentclass[12pt]{article}
\usepackage{amsfonts,amssymb,amsmath,graphicx}
\newtheorem{theorem}{Theorem}

\begin{document}
\parskip=3pt
\parindent=18pt
\baselineskip=20pt
\setcounter{page}{1}
\centerline{\large\bf Sum uncertainty relations based on Wigner-Yanase skew information}
\vspace{6ex}
\centerline{{\sf Bin Chen,$^{\star,\dag}$}
\footnote{\sf chenbin5134@163.com}
~~~ {\sf Shao-Ming Fei,$^{\natural,\sharp}$}
\footnote{\sf feishm@cnu.edu.cn}
~~~ {\sf Gui-Lu Long$^{\star,\dag,\ddag}$}
\footnote{\sf gllong@tsinghua.edu.cn}
}
\vspace{4ex}
\centerline
{\it $^\star$ State Key Laboratory of Low-Dimensional Quantum Physics and Department of Physics,}\par
\centerline
{\it Tsinghua University, Beijing 100084, China}\par
\centerline
{\it $^\dag$ Tsinghua National Laboratory for Information Science and Technology, Beijing 100084, China}\par
\centerline
{\it $^\natural$ School of Mathematical Sciences, Capital Normal University, Beijing 100048, China}\par
\centerline
{\it $^\sharp$ Max-Planck-Institute for Mathematics in the Sciences, 04103 Leipzig, Germany}\par
\centerline
{\it $^\ddag$ Collaborative Innovation Center of Quantum Matter, Beijing 100084, China}\par
\vspace{6.5ex}
\parindent=18pt
\parskip=5pt
\begin{center}
\begin{minipage}{5in}
\vspace{3ex}
\centerline{\large Abstract}
\vspace{4ex}
We study sum uncertainty relations for arbitrary finite $N$ quantum mechanical observables.
Some uncertainty inequalities are presented by using skew information introduced by Wigner and Yanase.
These uncertainty inequalities are nontrivial as long as the observables are mutually noncommutative.
The relations among these new and existing uncertainty inequalities have been investigated.
Detailed examples are presented.
\end{minipage}
\end{center}

\newpage

\section{Introduction}
Uncertainty principle is considered to be one of the most singular characteristics of quantum mechanics.
Ever since the first uncertainty relation of position-momentum was proposed by Heisenberg in 1927 \cite{Heisenberg}, many efforts have been made in interpreting the intrinsic meaning of this kind of inequalities.
The traditional approach that formulates uncertainty relations is based on the variance of measurement outcomes.
The most famous one is due to Robertson \cite{Robertson}, who derived the following uncertainty inequality for two observables $A$ and $B$:
\begin{equation}\label{UR}
\Delta_{\rho} A\Delta_{\rho} B\geq\frac{1}{2}|\langle[A,B]\rangle_{\rho}|,
\end{equation}
where $\Delta_{\rho}\Omega=\sqrt{\langle \Omega^{2}\rangle_{\rho}-\langle \Omega\rangle_{\rho}^{2}}$ is the standard deviation of an observable $\Omega$, and $[A,B]=AB-BA$.
It is obvious that the above relation is nontrivial for two non-commuting observables when suitable states are measured.
That is to say, the lower bound in (\ref{UR}) can be used to capture the non-commutativity of two observables.
There also have been many ways to describe uncertainty relations, such as in terms of entropies \cite{y10,y11,j1,j2,j3,j4}
and by means of majorization technique \cite{y12,y13,y131,y132,y14,y141}.

Based on several nice properties such as convexity and additivity, Wigner and Yanase (WY) introduced their skew information.
Being a measure for the non-commutativity between a state $\rho$ and an observable $H$, the skew information provides a measure of quantum uncertainty of $H$ in the state $\rho$, and
was used by Luo to derive a refinement of Heisenberg's uncertainty relation for mixed states \cite{Luo05}.

On the other hand, the skew information
$$
I_{\rho}(H)=-\frac{1}{2}\mathrm{Tr}([\sqrt{\rho},H]^{2})=\frac{1}{2}\|[\sqrt{\rho},H]\|^{2}
$$
was introduced by Wigner and Yanase in 1963 \cite{WY}
to quantify the information content of a quantum state $\rho$ with respect to the observables not commuting
with (i.e., skew to) the conserved quantity $H$.
It becomes a useful tool in quantum information theory in recent years,
such as characterizing nonclassical correlations \cite{Girolami}, being a measure of the $H$ coherence of the state $\rho$ \cite{Girolami2},
and quantifying the dynamics of some physical phenomena \cite{Metwally,Sun,Sun2}.

However, previous works demonstrated that the skew information is not suitable for formulating uncertainty relations: the following relation:
$$
I_{\rho}(A)I_{\rho}(B)\geq\frac{1}{4}|\langle[A,B]\rangle_{\rho}|^{2}
$$
turned to be false \cite{LuoZhang04,LuoZhang05,Kosaki,Yanagi}.
Alternatively, in Ref.\cite{Luo05} Luo defined the quantity $U_{\rho}(H)$,
$$
U_{\rho}(H)=\sqrt{(\Delta_{\rho} H)^{4}-[(\Delta_{\rho} H)^{2}-I_{\rho}(H)]^{2}},
$$
and provided a new Heisenberg-type uncertainty relation:
\begin{equation}\label{UUR}
U_{\rho}(A)U_{\rho}(B)\geq\frac{1}{4}|\langle[A,B]\rangle_{\rho}|^{2}.
\end{equation}
After that, Furuichi \cite{Furuichi} derived a Schr\"{o}dinger-type uncertainty relation using the quantity $U_{\rho}(H)$,
and a generalization of Schr\"{o}dinger's uncertainty relation based on Wigner-Yanase skew information is presented in Ref. \cite{caohuaixin}.

Besides uncertainty relations in the product form, variance-based and standard deviation-based sum uncertainty relations are also attracting considerable attention recently \cite{Maccone,Chen,Pati,Huang}.
In Ref. \cite{Maccone}, the authors derived two variance-based sum uncertainty relations for two incompatible observables.
They show that the lower bounds are nontrivial whenever the two observables are incompatible on the state of the system.
Thus the lower bounds of these uncertainty inequalities can be used to capture the incompatibility of the observables.
After that, Chen and Fei \cite{Chen} presented two uncertainty inequalities in terms of the sum of variances and standard deviations
for arbitrary $N$ incompatible observables, respectively.

In this paper, we formulate sum uncertainty relations by use of the skew information.
We presented several uncertainty inequalities for arbitrary $N$ observables.
These lower bounds have explicit physical meaning, i.e., they can be used to capture the non-commutativity
of the observables like Heisenberg-Robertson product uncertainty relation (\ref{UR}). Some applications and examples are also provided.

\section{Skew information-based sum uncertainty relations for two observables}
We first present two skew information-based sum uncertainty inequalities for two observables.
\begin{theorem}
For two observables $A$ and $B$, we have
\begin{equation}\label{sk1}
I_{\rho}(A)+I_{\rho}(B)\geq\frac{1}{2}\max\{I_{\rho}(A+B),I_{\rho}(A-B)\}.
\end{equation}
\end{theorem}

\emph{Proof.} Consider the parallelogram law in a Hilbert space: $\|u+v\|^{2}+\|u-v\|^{2}=2(\|u\|^{2}+\|v\|^{2})$.
Let $u=[\sqrt{\rho},A],v=[\sqrt{\rho},B]$. Then we have
\begin{equation}
\begin{split}
I_{\rho}(A)+I_{\rho}(B)&=\frac{1}{2}[I_{\rho}(A+B)+I_{\rho}(A-B)]\\
&\geq\frac{1}{2}\max\{I_{\rho}(A+B),I_{\rho}(A-B)\}.
\end{split}
\end{equation}
This completes the proof.  \quad $\Box$

If the lower bound in (\ref{sk1}) is zero, then we have $[\sqrt{\rho},A+B]=[\sqrt{\rho},A-B]=0$, which implies that $[\sqrt{\rho},A]=[\sqrt{\rho},B]=0$. Further, we can conclude that $[\rho,A]=[\rho,B]=0$ and $\langle[A,B]\rangle_{\rho}=0$.
Hence it is easily to see that if the two observables $A$ and $B$ are noncommutative, the relation (\ref{sk1}) is nontrivial.
That is to say, skew information can be used to formulate sum uncertainty relations, and the lower bound we derived in (\ref{sk1}) is meaningful.
In particular, for pure states we get the uncertainty relations presented in Ref. \cite{Maccone,Yao}, since in this case $I_{\rho}(H)=(\Delta_{\rho} H)^{2}$.

Similar to the standard deviation-based sum uncertainty relations, we provide another uncertainty inequality by using the quantity $\sqrt{I_{\rho}(H)}$.
From the inequality $\|u\pm v\|\leq\|u\|+\|v\|$, we have the following Theorem.
\begin{theorem}
For two observables $A$ and $B$, we have
\begin{equation}\label{sk2}
\sqrt{I_{\rho}(A)}+\sqrt{I_{\rho}(B)}\geq\max\{\sqrt{I_{\rho}(A+B)},\sqrt{I_{\rho}(A-B)}\}.
\end{equation}
\end{theorem}

It is obvious that the lower bound in (\ref{sk2}) is nonzero as long as $A$ and $B$ are noncommutative.
Thus the relation (\ref{sk2}) can be used to characterize the non-commutativity of $A$ and $B$ too.

\section{Skew information-based sum uncertainty relations for arbitrary $N$ observables}
Before generalizing the uncertainty relations (\ref{sk1}) and (\ref{sk2}) to arbitrary $N$ observables case,
let us first improve several uncertainty relations for $N$ observables.
Let $A_{1},\ldots,A_{N}$ be $N$ observables. Since
\begin{equation}
\begin{split}
\sqrt{I_{\rho}\left(\sum_{i=1}^{N}A_{i}\right)}
&=\frac{1}{\sqrt{2}}\left\|\left[\sqrt{\rho},\sum_{i=1}^{N}A_{i}\right]\right\|\\
&\leq\frac{1}{\sqrt{2}}\sum_{i=1}^{N}\|[\sqrt{\rho},A_{i}]\|\\
&=\sum_{i=1}^{N}\sqrt{I_{\rho}(A_{i})},
\end{split}
\end{equation}
we get
\begin{equation}\label{sq1}
\sum_{i=1}^{N}\sqrt{I_{\rho}(A_{i})}\geq\sqrt{I_{\rho}\left(\sum_{i=1}^{N}A_{i}\right)}.
\end{equation}
If $\rho$ is a pure state, then the relation (\ref{sq1}) gives rise to the result presented in Ref. \cite{Pati}:
$$
\sum_{i}\Delta_{\rho}(A_{i})\geq\Delta_{\rho}\left(\sum_{i}A_{i}\right).
$$

Let $\{A_{i}\}_{i=1}^{N}$ and $\{B_{j}\}_{j=1}^{N}$ be two sets of non-commuting observables satisfying $[A_{i},B_{j}]=\mathrm{i}\delta_{ij}C$.
In Ref. \cite{Pati} it has been proved that the product of sum of uncertainties in the individual observables satisfies the following inequality:
\begin{equation}\label{XP}
\left(\sum_{i=1}^{N}\Delta_{\rho} A_{i}\right)\left(\sum_{i=1}^{N}\Delta_{\rho} B_{j}\right)\geq\frac{N}{2}|\langle C\rangle_{\rho}|.
\end{equation}
By using uncertainty relation (\ref{UUR}) and noting that $\sqrt{U_{\rho}(H)}\leq\Delta_{\rho}H$, we have a refined form of (\ref{XP}):
\begin{equation}\label{UXP}
\begin{split}
\left(\sum_{i=1}^{N}\sqrt{U_{\rho}(A_{i})}\right)\left(\sum_{j=1}^{N}\sqrt{U_{\rho}(B_{j})}\right)&=\sum_{ij}\sqrt{U_{\rho}(A_{i})U_{\rho}(B_{j})}\\
&\geq\frac{1}{2}\sum_{ij}|\langle[A_{i},B_{j}]\rangle_{\rho}|\\
&=\frac{N}{2}|\langle C\rangle_{\rho}|.
\end{split}
\end{equation}
The above uncertainty relation (\ref{XP}) can be further improved:
\begin{theorem}
Let $\{A_{i}\}_{i=1}^{N}$ and $\{B_{j}\}_{j=1}^{N}$ be two sets of non-commuting observables satisfying $[A_{i},B_{j}]=\mathrm{i}\delta_{ij}C$. Then we have
\begin{equation}\label{RUXP}
\left(\sum_{i=1}^{N}U_{\rho}(A_{i})\right)\left(\sum_{j=1}^{N}U_{\rho}(B_{j})\right)\geq\frac{N}{4}|\langle C\rangle_{\rho}|^{2}.
\end{equation}
\end{theorem}

\emph{Proof.}
\begin{equation}
\begin{split}
\left(\sum_{i=1}^{N}U_{\rho}(A_{i})\right)\left(\sum_{j=1}^{N}U_{\rho}(B_{j})\right)&=\sum_{ij}U_{\rho}(A_{i})U_{\rho}(B_{j})\\
&\geq\frac{1}{4}\sum_{ij}|\langle[A_{i},B_{j}]\rangle_{\rho}|^{2}\\
&=\frac{N}{4}|\langle C\rangle_{\rho}|^{2}.
\end{split}
\end{equation}
To show that the uncertainty relation (\ref{RUXP}) is stronger than (\ref{UXP}), one only need to note that $\left(\sum_{i=1}^{N}\sqrt{U_{\rho}(\Omega_{i})}\right)^{2}\geq\left(\sum_{i=1}^{N}U_{\rho}(\Omega_{i})\right)$, $\Omega=A,B$.  \quad $\Box$

Let us now generalize the uncertainty relation (\ref{sk1}) to arbitrary $N$ observables.
\begin{theorem}
For arbitrary $N$ observables $A_{1}$, $A_{2}$, $\ldots$, $A_{N}$, we have
\begin{equation}\label{Nsk1}
\sum_{i=1}^{N}I_{\rho}(A_{i})\geq\frac{1}{N-2}\left[\sum_{1\leq i<j\leq N}I_{\rho}(A_{i}+A_{j})
-\frac{1}{(N-1)^{2}}\left(\sum_{1\leq i<j\leq N}\sqrt{I_{\rho}(A_{i}+A_{j})}\right)^{2}\right].
\end{equation}
If $A_{i}s$ are mutually noncommutative, then the lower bound in (\ref{Nsk1}) is nonzero.
\end{theorem}

\emph{Proof.}
Consider the following inequality in a Hilbert space \cite{Chen}:
\begin{equation}\label{13}
\begin{split}
\sum_{i=1}^{N}\|u_{i}\|^{2}&\geq\frac{1}{N-2}\left[\sum_{1\leq i<j\leq N}\|u_{i}+u_{j}\|^{2}
-\frac{1}{(N-1)^{2}}\left(\sum_{1\leq i<j\leq N}\|u_{i}+u_{j}\|\right)^{2}\right]\\
&\geq\frac{1}{2(N-1)}\sum_{1\leq i<j\leq N}\|u_{i}+u_{j}\|^{2}.
\end{split}
\end{equation}
Let $u_{i}=[\sqrt{\rho},A_{i}]$. We obtain (\ref{Nsk1}) directly. Moreover, if
the lower bound (\ref{Nsk1}) is zero, then each $I_{\rho}(A_{i}+A_{j})$ is equal to zero, which implies that $I_{\rho}(A_{i})=0,\forall i$. Thus we have $\langle[A_{i},A_{j}]\rangle_{\rho}=0$. That is to say, for $N$ non-commuting observables, the relation (\ref{Nsk1}) is nontrivial.  \quad $\Box$

{\it Remark}. From (\ref{sk1}) one may also trivially derive an uncertainty relation for $N$ observables by adding the skew information of every pair of the
observables. Nevertheless, it can been shown from (\ref{13}) that the lower bound in (\ref{Nsk1}) is tighter than the one trivially derived from (\ref{sk1}) in such way.

We now generalize uncertainty relation (\ref{sk2}) to arbitrary $N$ observables case.
\begin{theorem}
For arbitrary $N$ observables $A_{1}$, $A_{2}$, $\ldots$, $A_{N}$, we have
\begin{equation}\label{Nsk2}
\sum_{i=1}^{N}\sqrt{I_{\rho}(A_{i})}\geq\frac{1}{N-2}\left[\sum_{1\leq i<j\leq N}\sqrt{I_{\rho}(A_{i}+A_{j})}
-\sqrt{I_{\rho}\left(\sum_{i=1}^{N}A_{i}\right)}\right].
\end{equation}
If $A_{i}s$ are mutually noncommutative, then the lower bound in (\ref{Nsk2}) is nonzero.
\end{theorem}

\emph{Proof.} Using the following inequality \cite{Chen,Horn,Honda},
\begin{equation}\label{15}
\begin{split}
\sum_{i=1}^{N}\|u_{i}\|&\geq\frac{1}{N-2}\left(\sum_{1\leq i<j\leq N}\|u_{i}+u_{j}\|-\left\|\sum_{i=1}^{N}u_{i}\right\|\right)\\
&\geq\left\|\sum_{i=1}^{N}u_{i}\right\|,
\end{split}
\end{equation}
and setting $u_{i}=[\sqrt{\rho},A_{i}]$, we get (\ref{Nsk2}). If the lower bound in (\ref{Nsk2}) is zero,
so is (\ref{sq1}), and all $I_{\rho}(A_{i}+A_{j})$ are equal to zero. In this case we have $I_{\rho}(A_{i})=0,\forall i$,
and $\langle[A_{i},A_{j}]\rangle_{\rho}=0$. Thus as long as $A_{i}s$ are mutually noncommutative, the relation (\ref{Nsk2}) is nontrivial.  \quad $\Box$

It can be verified from (\ref{15}) that the inequality (\ref{Nsk2}) is tighter than (\ref{sq1}).
In the following, we provide a new skew information-based sum uncertainty relation.
\begin{theorem}
Let $A_{1}$, $A_{2}$, $\ldots$, $A_{N}$ be $N$ observables, $\rho$ a quantum state such that $I_{\rho}(A_{i})\neq0, \forall i$. Let $G$ be an $N\times N$ matrix with entries $G_{ij}=\mathrm{Tr}(X_{i}X_{j})$, where\\
$X_{i}=\mathrm{i}[\sqrt{\rho},A_{i}]/\|[\sqrt{\rho},A_{i}]\|, \mathrm{i}=\sqrt{-1}, \forall i$. Then we have
\begin{equation}\label{SNsk2}
\sum_{i=1}^{N}I_{\rho}(A_{i})\geq\frac{1}{\lambda_{\max}(G)}I_{\rho}\left(\sum_{i=1}^{N}A_{i}\right),
\end{equation}
where $\lambda_{\max}(G)$ denotes the maximal eigenvalue of $G$.
\end{theorem}

\emph{Proof.} Since $X_{i}s$ are Hermitian matrices, $G$ is also Hermitian. Further, $G$ is a positive semi-definite matrix, since
$\sum_{i,j}x_{i}^{*}G_{ij}x_{j}=\sum_{i,j}x_{i}^{*}\mathrm{Tr}(X_{i}X_{j})x_{j}=\|\sum_{i}x_{i}X_{i}\|^{2}\geq0$.
Thus all eigenvalues of $G$ are nonnegative numbers. Note that
\begin{equation}
\begin{split}
I_{\rho}\left(\sum_{i}A_{i}\right)&=-\frac{1}{2}\mathrm{Tr}\left(\left[\sqrt{\rho},\sum_{i}A_{i}\right]^{2}\right)\\
&=\frac{1}{2}\sum_{i,j}\mathrm{Tr}\left[(\mathrm{i}[\sqrt{\rho},A_{i}])(\mathrm{i}[\sqrt{\rho},A_{j}])\right]\\
&=\frac{1}{2}\sum_{i,j}\|[\sqrt{\rho},A_{i}]\|G_{ij}\|[\sqrt{\rho},A_{j}]\|\\
&=\sum_{i,j}\sqrt{I_{\rho}(A_{i})}G_{ij}\sqrt{I_{\rho}(A_{j})}\\
&\leq\lambda_{\max}(G)\sum_{i}I_{\rho}(A_{i}),
\end{split}
\end{equation}
we get (\ref{SNsk2}).  \quad $\Box$

\begin{figure}
\centering
\includegraphics[width=7cm]{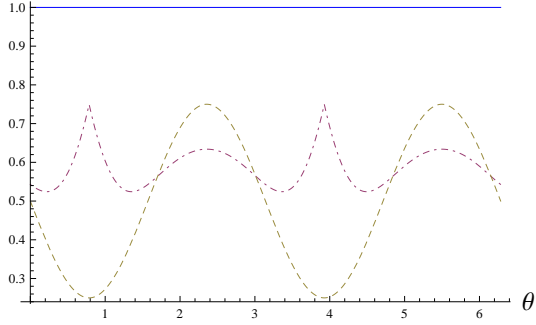}
\caption{The horizontal line is the sum of the skew information $I_{\rho}(\sigma_{1})+I_{\rho}(\sigma_{2})+I_{\rho}(\sigma_{3})$.
The dot-dashed line is the bound (\ref{Nsk1}).The dashed line is the bound (\ref{SNsk2}).}\label{chenb1}
\end{figure}

As an example, let us consider three observables case, the Pauli matrices
$\sigma_{1}=|0\rangle\langle1|+|1\rangle\langle0|$, $\sigma_{2}=-i|0\rangle\langle1|+i|1\rangle\langle0|$ and $\sigma_{3}=|0\rangle\langle0|-|1\rangle\langle1|$.
Let the measured states be given with Bloch vector $\overrightarrow{r}=(\frac{\sqrt{3}}{2}\cos\theta,\frac{\sqrt{3}}{2}\sin\theta,0)$.
Then we have
$I_{\rho}(\sigma_{1})=\frac{1}{2}\sin^{2}\theta$, $I_{\rho}(\sigma_{2})=\frac{1}{2}\cos^{2}\theta$, $I_{\rho}(\sigma_{3})=\frac{1}{2}$,
$\lambda_{\max}(G)=2$, $I_{\rho}(\sigma_{1}+\sigma_{2})=\frac{1}{2}(1-\sin2\theta)$, $I_{\rho}(\sigma_{2}+\sigma_{3})=\frac{1}{4}(3+\cos2\theta)$,
$I_{\rho}(\sigma_{1}+\sigma_{3})=\frac{1}{4}(3-\cos2\theta)$, $I_{\rho}(\sigma_{1}+\sigma_{2}+\sigma_{3})=1-\frac{1}{2}\sin2\theta$.
The comparison between the lower bounds (\ref{Nsk1}) and (\ref{SNsk2}) is given in FIG \ref{chenb1}.

One can see that (\ref{SNsk2}) is stronger than (\ref{Nsk1}) in some cases.
Therefore, we have the following skew information-based sum uncertainty relation:
\begin{theorem}
\begin{equation}
\sum_{i=1}^{N}I_{\rho}(A_{i})\geq\max\{(lb1),(lb2)\},
\end{equation}
where $(lb1)$ and $(lb2)$ stand for the lower bounds in (\ref{Nsk1}) and (\ref{SNsk2}), respectively.
\end{theorem}

\section{Conclusion}
Skew information is an important quantity in the theory of quantum information,
and plays significant roles in many quantum information processing tasks.
In this paper, we have demonstrated that skew information can be used to formulate sum uncertainty relations for arbitrary $N$ observables. 
The skew information-based sum uncertainty relations are rather different from the variance-based ones like in Ref. \cite{Maccone,Chen,Yao}.
First, the physical meaning of skew information is quite different to the variances of measurement outcomes, although they are numerically equal when the physical system is in a pure quantum state.
On the other hand, uncertainty relations based on skew information and variances have their own advantages.
The lower bounds of variance-based sum uncertainty relations in Ref. \cite{Maccone,Chen,Yao} capture the incompatibility of the observables, i.e., the lower bound is nonzero as long as the quantum states are not the common eigenstates of the observables.
While the lower bounds of skew information-based uncertainty relations we derived in this paper capture the non-commutativity of the observables, i.e., the inequalities are nontrivial when the observables are noncommutative with respect to the measured states. 
For pure states, some of these skew information-based inequalities are reduced to the variance-based ones \cite{Maccone,Chen,Yao}.
We hope that these results may shed lights on further investigations on skew information-based uncertainty relations.

\vspace{2.5ex}
\noindent{\bf Acknowledgments}\, \,
We would like to thank Dr. Shu-Hao Wang for his helpful discussions and suggestions.
This work is supported by the National Basic Research Program of China (2015CB921002),
the National Natural Science Foundation of China Grant Nos. 11175094, 91221205 and 11275131.


\begin{thebibliography}{18}

\bibitem{Heisenberg}
Heisenberg, W.: \"{U}ber den anschaulichen Inhalt der quantentheoretischen Kinematik und Mechanik. Z. Phys. \textbf{43}, 172 (1927)

\bibitem{Robertson}
Robertson, H. P.: The uncertainty principle. Phys. Rev. \textbf{34}, 163¨C164 (1929)

\bibitem{y10}
Deutsch, D.: Uncertainty in Quantum Measurements. Phys. Rev. Lett. \textbf{50}, 631 (1983)

\bibitem{y11}
Maassen, H., Uffink, J. B. M.: Generalized entropic uncertainty relations. Phys. Rev. Lett. \textbf{60}, 1103 (1988)

\bibitem{j1}
Wehner, S., Winter, A.: Entropic uncertainty relations -- a survey. New J. Phys. \textbf{12}, 025009 (2010)

\bibitem{j2}
Wu, S., Yu, S., M{\o}lmer, K.: Entropic uncertainty relation for mutually unbiased bases. Phys. Rev. A \textbf{79}, 022104 (2009)

\bibitem{j3}
Rastegin, A. E.: Uncertainty relations for MUBs and SIC-POVMs in terms of generalized entropies. Eur. Phys. J. D \textbf{67}, 269 (2013)

\bibitem{j4}
Chen, B., Fei, S. M.: Uncertainty relations based on mutually unbiased measurements. Quantum Inf. Process. \textbf{14}, 2227 (2015)

\bibitem{y12}
Bialynicki-Birula, I., Rudnicki, L.: Entropic Uncertainty Relations in Quantum Physics.
In: Statistical Complexity. Sen, K. D. (eds.). Springer, Berlin (2011)

\bibitem{y13}
Puchala, Z., Rudnicki, L., Zyczkowski, K.: Majorization entropic uncertainty relations. J. Phys. A: Math. Theor. \textbf{46}, 272002 (2013)

\bibitem{y131}
Rudnicki, L., Puchala, Z., Zyczkowski, K.: Strong majorization entropic uncertainty relations. Phys. Rev. A \textbf{89}, 052115 (2014)

\bibitem{y132}
Rudnicki, L.: Majorization approach to entropic uncertainty relations for coarse-grained observables. Phys. Rev. A \textbf{91}, 032123 (2015)

\bibitem{y14}
Friedland, S., Gheorghiu, V., Gour, G.: Universal Uncertainty Relations. Phys. Rev. Lett. \textbf{111}, 230401 (2013)

\bibitem{y141}
Narasimhachar, V., Poostindouz, A., Gour, G.: The principle behind the Uncertainty Principle. arXiv:1505.02223

\bibitem{WY}
Wigner, E. P., Yanase, M. M.: Information contents of distributions. Proc. Natl. Acad. Sci. U.S.A. \textbf{49}, 910 (1963)

\bibitem{Luo05}
Luo, S.: Heisenberg uncertainty relation for mixed states. Phys. Rev. A \textbf{72}, 042110 (2005)

\bibitem{LuoZhang04}
Luo, S., Zhang, Q.: On skew information. IEEE Trans. Inf. Theory \textbf{50}, 1778 (2004)

\bibitem{LuoZhang05}
Luo, S., Zhang, Q.: Correction to "On Skew Information". IEEE Trans. Inf. Theory \textbf{51}, 4432 (2005)

\bibitem{Kosaki}
Kosaki, H.: Matrix trace inequalities related to uncertainty principle. Int. J. Math. \textbf{16}, 629 (2005)

\bibitem{Yanagi}
Yanagi, K., Furuichi, S., Kuriyama, K.: A generalized skew information and uncertainty relation. IEEE Trans. Inf. Theory \textbf{51}, 4401 (2005)

\bibitem{Girolami}
Girolami, D., Tufarelli, T., Adesso, G.: Characterizing nonclassical correlations via local quantum uncertainty. Phys. Rev. Lett. \textbf{110}, 240402 (2013)

\bibitem{Girolami2}
Girolami, D.: Observable measure of quantum coherence in finite dimensional systems. Phys. Rev. Lett. \textbf{113}, 170401 (2014)

\bibitem{Metwally}
Metwally, N., Al-Mannai, A., Abdel-Aty, M.: Skew Information for a Single Cooper Pair Box Interacting with a Single Cavity Field. Commun. Theor. Phys. \textbf{59}, 769 (2013)

\bibitem{Sun}
Sun, H. G., Liu, W. F., Li, C. J.: Maximal and total skew information for a two-qubit system using nonlinear interaction models. Chin. Phys. B \textbf{20}, 090301 (2011)

\bibitem{Sun2}
Sun, H. G., Zhang, L. H., Liu, W. F., Li, C. J.: Maximal and total skew information of three-qubit system obtained using nonlinear interaction models. Chin. Phys. B \textbf{21}, 010301 (2012)

\bibitem{Furuichi}
Furuichi, S.: Schr\"{o}dinger uncertainty relation with Wigner-Yanase skew information. Phys. Rev. A \textbf{82}, 034101 (2010)

\bibitem{caohuaixin}
Li, Q., Cao, H. X., Du, H. K.: A generalization of Schr\"{o}dinger's uncertainty relation described by the Wigner-Yanase skew information. Quantum. Inf. Process. \textbf{14}, 1513 (2015)

\bibitem{Maccone}
Maccone, L., Pati, A. K.: Stronger uncertainty relations for all incompatible observables. Phys. Rev. Lett. \textbf{113}, 260401 (2014)

\bibitem{Chen}
Chen, B., Fei, S. M.: Sum uncertainty relations for arbitrary $N$ incompatible observables. Scientific Reports \textbf{5}, 14238 (2015)

\bibitem{Pati}
Pati, A. K., Sahu, P. K.: Sum uncertainty relation in quantum theory. Physics Letters A \textbf{367}, 177 (2007)

\bibitem{Huang}
Huang, Y.: Variance-based uncertainty relations. Phys. Rev. A \textbf{86}, 024101 (2012)

\bibitem{Yao}
Yao, Y., Xiao, X., Wang, X., Sun, C. P.: Implications and applications of the variance-based uncertainty equalities. Phys. Rev. A \textbf{91}, 062113 (2015)

\bibitem{Horn}
Horn, R. A., Johnson, C. R.: Matrix Analysis. 2nd ed. Cambridge University Press, Cambridge, England (2013)

\bibitem{Honda}
Honda, A., Okazaki, Y., Takahashi, Y.: Generalizations of the Hlawka's inequality. Pure Appl. Math. \textbf{45}, 9-15 (1998)







\end{thebibliography}
\end{document}